\documentclass[9pt,twocolumn,twoside]{osajnl_teemu}

\usepackage{epstopdf}

\journal{ol} 

\setboolean{shortarticle}{true} 

\ifthenelse{\boolean{shortarticle}}{\colorlet{color2}{color2b}}{\colorlet{color2}{color2}} 

\title{Sub-part-per-trillion level detection of hydrogen fluoride by cantilever-enhanced photo-acoustic spectroscopy}

\author[1]{Teemu Tomberg}
\author[1,2]{Markku Vainio}
\author[3]{Tuomas Hieta}
\author[1,*]{Lauri Halonen}

\affil[1]{Department of Chemistry, University of Helsinki, P.O. Box 55, FIN-00014 Helsinki, Finland}
\affil[2]{VTT Technical Research Centre of Finland Ltd., Centre of Metrology MIKES, P.O. Box 1000, FI-02044 VTT, Finland}
\affil[3]{Gasera Ltd., Lemminkäisenkatu 59, FIN-20520 Turku, Finland}

\affil[*]{Corresponding author: lauri.halonen@helsinki.fi}

\dates{Compiled \today}

\ociscodes{(120.0120) Instrumentation, measurement, and metrology; (300.6430) Spectroscopy, photothermal;
(300.6340) Spectroscopy, infrared; (190.4970) Parametric oscillators and amplifiers.}

\doi{}

\begin{abstract}
We present a highly sensitive hydrogen fluoride detector based on simple cantilever-enhanced photo-acoustic spectroscopy. A noise equivalent concentration of 5 parts-per-trillion was achieved in 1~s and 650 parts-per-quadrillion in 32 min. The limits were reached with an average optical power of 950~mW using a new continuous-wave optical parametric oscillator at 2.476~\textmu m. The achieved normalized noise equivalent absorption coefficient was $2.7\times 10^{-10}$~W\,cm$^{-1}$Hz$^{-1/2}$.
\end{abstract}

\setboolean{displaycopyright}{false}

\begin{document}

\maketitle

\ifthenelse{\boolean{shortarticle}}{\ifthenelse{\boolean{singlecolumn}}{\abscontentformatted}{\abscontent}}{}


\noindent Hydrogen fluoride (HF) is a highly reactive and corrosive gas that is used widely in industrial processes, such as etching and in the production of refrigerants, high-octane gasoline and aluminium. It is hazardous to health when in contact with skin or inhaled, having a permissible exposure limit of 3 parts-per-million-by-volume (ppm$_\textrm{v}$) \cite{NIOSH2007}. In the troposphere, the short lifetime of HF assures safe overall concentration, estimated to be less than a few parts-per-trillion (ppt$_\textrm{v}$) \cite{Duchatelet2010}. Hydrogen fluoride enters the troposphere mainly from anthropogenic sources, such as coal-fired power stations and aluminium smelters. Natural sources of fluorides, such as volcano eruptions, geysers and marine aerosols, make only a small contribution to the total emission rates \cite{Walna2013}. Near emission sources, HF concentrations may rise to a few tens of parts-per-billion (ppb$_\textrm{v}$), which already has harmful effects on vegetation \cite{Walna2013}. To detect and prevent any harmful leaks or accidents, even the smallest concentrations must be monitored. Mixing ratios of a few ppm$_\textrm{v}$ are conveniently measured using various chemical reaction-based methods \cite{Lim2009,Mertens2004}, whereas spectroscopic techniques are better suited for ppb$_\textrm{v}$ level detection \cite{Dressler2008,Frish2005}. Recently, a noise equivalent HF concentration of 38 ppt$_\textrm{v}$ in 1~s was reported by Craig et al. using tunable diode laser absorption spectroscopy \cite{Craig2015}.

Spectroscopic techniques are promising in further improving the detection limit of HF. Hydrogen fluoride has a strong rotational-vibrational absorption band in the near-infrared region between 2.3 and 2.8~\textmu m \cite{Rothman2013}, which enables a high sensitivity \emph{in situ} detection using laser-based absorption spectroscopic methods \cite{Tittel2003,Vainio2016}. One of the most sensitive spectroscopic techniques is cantilever-enhanced photo-acoustic spectroscopy (CEPAS), which has been shown to reach a noise equivalent detection limit as low as 50~ppt in the case of NO$_2$ \cite{Peltola2015}.

Photo-acoustic spectroscopy (PAS) is a wavelength-independent technique for absorption spectroscopy. In PAS, a photo-acoustic (PA) signal is produced by non-radiative relaxation of periodically excited molecules, conventionally detected by a microphone. The PA signal is directly proportional to the exciting optical power, which favours the use of high-power laser sources. Furthermore, in recent years, the developments in PAS research have surpassed the limitations of the insensitive microphone with an extremely sensitive miniature silicon cantilever \cite{Koskinen2006,Koskinen2007}. In CEPAS, the PA signal is detected by measuring the movement of a silicon cantilever by a laser interferometer. The modulation frequency can be chosen freely below the resonance frequency of the cantilever, typically designed to around 600 Hz. The best signal-to-noise ratio (SNR) is usually achieved around 10 to 100~Hz with a commercially available CEPAS instrument \cite{Koskinen2007}. 

In this Letter, we demonstrate a method capable of detecting HF down to 650 parts-per-quadrillion-by-volume (ppq$_\textrm{v}$). To our knowledge, this is the lowest HF sensitivity ever reported by a laser-based technique and one of the few laser-based gas analysers capable of sub-ppt$_\textrm{v}$ level trace gas detection \cite{Galli2011,Galli2016}. The experimental method is based on sensitive cantilever-enhanced photo-acoustic detection using a new narrow-linewidth high-power optical parametric oscillator (OPO) light source, operating on a strong HF transition at 2.476~\textmu m. 



\begin{figure}[tbp]
	\centering
	\includegraphics[width=\linewidth]{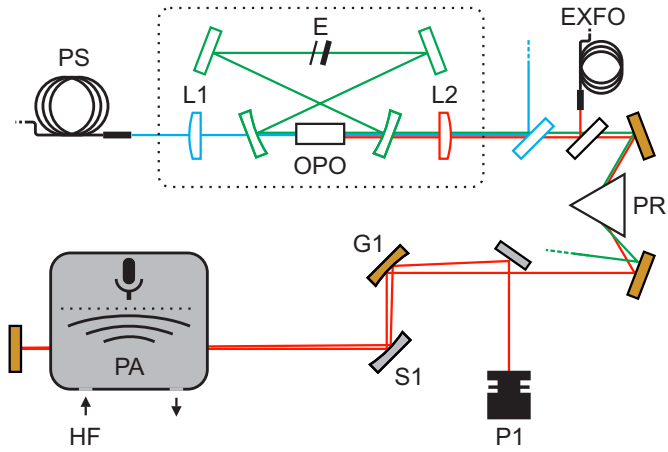}
	\caption{Schematic illustration of the experimental configuration. PS: pump laser source, L1: lens with \textit{f}=200~mm, L2: lens with \textit{f}=150~mm, E: two etalons, EXFO: wavemeter, PR: equilateral dispersive prism, G1: gold coated mirror with \textit{R}=200~mm, S1: silver coated mirror with \textit{R}=50~mm, P1: power meter, PA: photo-acoustic analyser, HF: hydrogen fluoride gas in dry air.}
	\label{fig:ExperimentalSetup}
\end{figure}

The experimental configuration of the HF detection system is shown schematically in Fig. \ref{fig:ExperimentalSetup}. The light source is a high-power optical parametric oscillator, producing coherent light on three principle wavelengths: 1064~nm pump, 1866~nm signal and 2476~nm idler. The idler beam is separated from the residual pump and signal beams using a combination of a dichroic mirror and an equilateral dispersive prism (Thorlabs PS853). In between, about 4\% of the idler and signal beams is sampled by a wedged uncoated CaF$_2$ window to an EXFO WA-1500 wavemeter for monitoring purposes. After the prism, the idler beam is guided through a cell of a PA analyzer in a double-pass configuration. The optical power after the cell is measured with a power meter. The photo-acoustic analyzer platform, manufactured by Gasera Ltd., is equipped with a cantilever-based photo-acoustic detection system. The PA cell has a length of 95~mm, a diameter of 4~mm, and total volume of about 7~ml. Its surface is gold coated. The windows are made of CaF$_2$ (Thorlabs WG50530-D) with a broad-band anti-reflection coating for the 2.47~\textmu m wavelength. The pressure in the cell is regulated to 200~mbar and the temperature to 50~$^{\circ}$C by an automatic control system. The system is configured to provide PA signal readings via a USB data link with an acquisition time of 1.0~s. The periodic 30~Hz photo-acoustic signal is formed by sinusoidal wavelength modulation of the OPO pump laser with 1.1~GHz modulation amplitude. A lock-in detection scheme is employed to perform a second harmonic signal detection. The modulation frequency was selected by looking at the acoustic noise spectrum of the CEPAS sensor and identifying interference free regions. 

\begin{figure}[htbp]
	\centering
	\includegraphics[width=\linewidth]{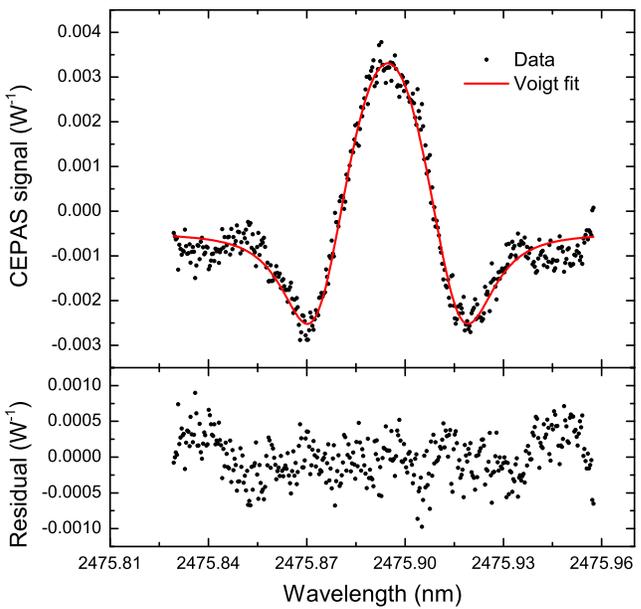}
	\caption{The upper figure shows a second harmonic signal of 97~ppt$_\textrm{v}$ HF in dry air, measured with an optical power of 740 mW, and a least squares fit of the second harmonic signal of a Voigt profile to the data. The lower figure shows the residual after subtraction of the fit.} 
	\label{fig:Spectrum}
\end{figure}

The design principle of the singly resonant continuous-wave OPO is described in previous articles \cite{Vainio2008a,Vainio2008b,Peltola2013}. For the new experiments reported here, the OPO had to be optimized to operate in a difficult wavelength region close to degeneracy. In addition, the long term stability was improved by almost a factor of 100 as compared to the previously reported OPOs \citep{Peltola2013}, which was crucial in terms of reaching the exceptional trace gas sensitivity via long averaging times. The OPO is pumped by a continuous-wave (CW) distributed feedback (DFB) laser (Eagleyard DFB:1064-0040-BFY02-0002) operating at 1064~nm and amplified by an ytterbium fiber amplifier (IPG YAR-20k-1064-LP-SF) to maximum optical output power of 20~W. The pump beam, after beam expansion, is focused into the OPO cavity with an anti-reflection coated lens. In the OPO, the pump beam is focused into a 50~mm-long MgO-doped, periodically poled lithium niobate (MgO:PPLN by HC Photonics) crystal, the poling period of which ranges from 26.5 to 32.5~\textmu m in a fanout pattern. The crystal is placed on an aluminium holder, the temperature of which can be stabilized anywhere between 20 and 100~$^{\circ}$C with a precision of $100$~mK using thermo-electric coolers and a commercial controller (Newport 350B). A bow-tie ring cavity is built around the crystal. The two plano-concave dielectric mirrors have a 125~mm radius of curvature, high transmission coatings for the pump and idler wavelength and a high reflectance for the signal wavelength. The two plane mirrors of the cavity have the same coatings. The cavity has a free-spectral-range (FSR) of 390~MHz. The relatively low FSR, operation near degeneracy and the strong water absorption around the resonant signal wavelength make the OPO spectrally and power-wise unstable without further measures. To solve this, the OPO cavity was placed in an enclosure that is continuously purged with dry air to reduce the humidity of the cavity to about one tenth of normal ambient humidity. This reduced the observed power fluctuations and increased the output power as the cavity losses were decreased. In addition, two uncoated YAG etalons (0.3~mm and 2~mm thick) were placed inside the cavity at the secondary focal point to narrow the OPO gain bandwidth and to reduce the intra cavity power levels by increasing the oscillation threshold \cite{Phillips2010}. The employed methods allowed us to stabilize the signal wavelength at a transmission maximum of water spectrum near 1866~nm, and to achieve highly stable single mode operation of more than three hours with around 50\% conversion efficiency, 980~mW idler output power and a linewidth less than 10~MHz. For lower power levels in the range of 100 to 600 mW, single mode operation up to several hours was achieved.

Photo-acoustic spectroscopy was performed on a strong rotational-vibrational transition of HF centered at 2475.8836~nm with a line intensity of 2.381$\times 10^{-18}$ cm$^{-1}$/(molecules cm$^{-2}$) at 296~K \cite{Rothman2013}. The wavelength in question is almost free of spectral interference from other atmospheric gases. The most significant interferer is a water line centered at 2475.72~nm. Simulations using the HITRAN 2012 database \cite{Rothman2013} demonstrated that the contribution of typical ambient water concentration of 1.4~\% at 200~mbar pressure creates a 30 times stronger second harmonic background signal than HF at a concentration of 1~ppt$_\textrm{v}$. A solution is to reduce the pressure, stabilize and closely monitor the water concentration, or reduce it to an insignificant level. Another reason to monitor the humidity is that adsorption of HF and H$_2$O in the measurement system causes the HF concentration to depend linearly on small changes of H$_2$O in a short time scale. 


\begin{figure}[tbp]
	\centering
	\includegraphics[width=\linewidth]{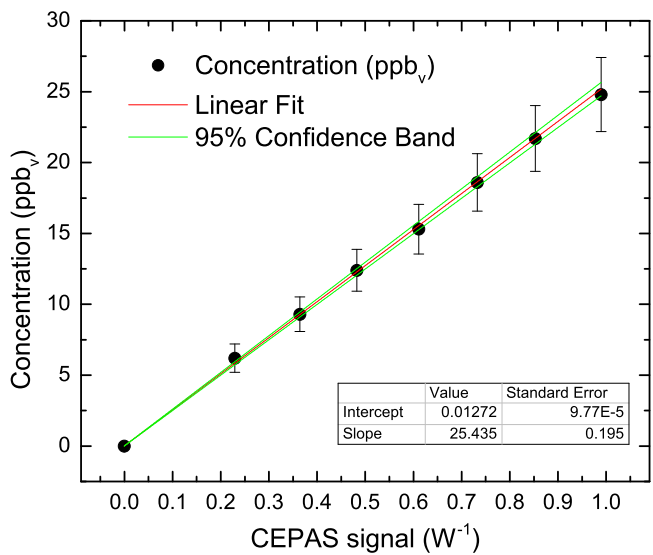}
	\caption{CEPAS signal as a function of HF concentration and a linear fit to the data. The measurement was performed using an optical power of 740~mW. Uncertainty of the MFCs and the concentration of the sample bottle is included.}
	\label{fig:Linearity}
\end{figure}

The performance of the CEPAS system was studied using the following gas analysis experiment. Two mass flow controllers (MFCs, \emph{Aera} and \emph{MKS Instruments}) were used to prepare a sample gas of desired HF concentration by mixing 3.05~ppm$_\textrm{v}$ of HF (AGA) and dry ambient air (humidity around 1050~ppm$_\textrm{v}$). Flow rate ranges were 4-200 and 40-2000~sccm for the MFC controlling HF and air, respectively. The controlled HF mixing ratios were then in the range of 6 to 400~ppb$_\textrm{v}$. Lower stable concentrations on the order of 100~ppt$_\textrm{v}$ were enabled by the slow desorption of HF in the sampling system while no HF was fed to the gas flow. The measurement cycle was 15~s long. 9~s of it was used for gas exchange by an automated gas exchange system and 6~s to record the PA signal at a 1~s sample rate. The set of six samples was then averaged to form one data point. Most measurements were carried out using an optical power in the range of 610 to 740~mW (instead of the maximum 980~mW) to ensure more consistent measurements as the OPO remained stable for longer periods.


\begin{figure}[hbt]
	\centering
	\includegraphics[width=\linewidth]{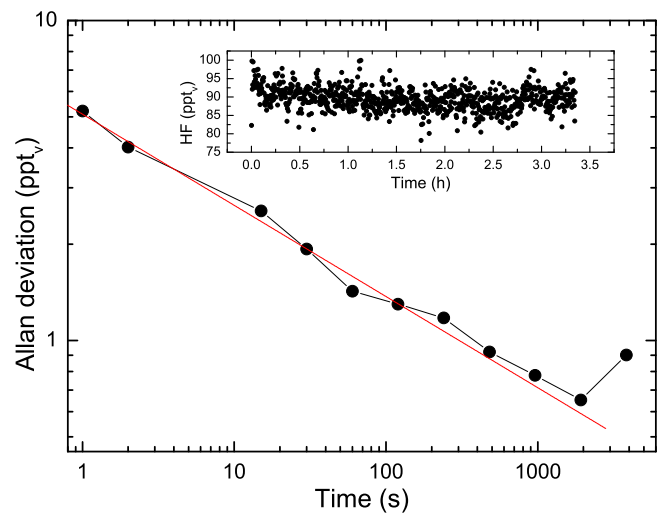}
	\caption{ Allan deviation of the HF volume mixing ratio as a function of averaging time. The inset shows the measurement data, obtained with optical power of 950 mW. The slope of the Allan deviation is -0.28.}
	\label{fig:AllanPlot}
\end{figure}


The experimental setup determines the HF concentration by measuring the peak second harmonic signal of the HF absorption line. In order to identify the HF line and possible background signals, the full spectrum is occasionally recorded as well. Figure \ref{fig:Spectrum} shows a recorded second harmonic sepctrum of 97~ppt$_\textrm{v}$ of HF. Black dots represent the measurement data recorded in about 14 minutes. A least squares second harmonic Voigt fit \cite{Westberg2012} (red curve) and its residual (lower figure) show good fit to the data. The obtained line parameters, including Lorentzian linewidth and modulation amplitude (MA), agree well with expectations. The modulation amplitude was included as a fit parameter to verify its value, which was adjusted to minimise residual amplitude modulation (RAM) in the second harmonic signal. The RAM was caused by an etalon effect in the PPLN crystal with non-optimal anti-reflection coatings for the idler beam. The RAM couples to the CEPAS signal through absorption in the windows of the PA cell. An estimate of the optimal modulation amplitude was derived from the measured FSR of the etalon and simulations on for which MA, relative to the FSR, the RAM in the second harmonic signal would go to zero \cite{DuBurck2004,Sun1992}. The RAM was reduced to an amplitude level equal to an HF signal of 8~ppt$_\textrm{v}$, as seen on the tails of the residual in Fig. \ref{fig:Spectrum}. The improvement was more than ten-fold, eliminating RAM from the performance limiting factors of the system as the residual RAM was found to be stable and predictable.


Photo-acoustic analyzers always require calibration. The data points for the calibration were collected in two steps. First, the power normalized CEPAS signal was measured at different controlled HF concentrations. Second, the zero HF point was determined as the background signal level of a spectral scan in Fig. \ref{fig:Spectrum}. A linear fit to the data, weighted by the uncertainty of the data points, was then determined as the response curve of the CEPAS system. The result, shown in Fig. \ref{fig:Linearity}, confirms a linear response of the system. The observed offset is caused by interfering absorbance of a nearby water transition, as expected by the simulations. Uncertainty in the produced analyte concentration, illustrated by the error bars, take into account the accuracy of the MFCs and the uncertainty of the sample bottle concentration. Another source of measurement error was adsorption of HF, which has to be carefully acknowledged for reproducible measurements. In the case of highly adsorptive molecules, such as HF, it is easier to remove than to enrich molecules on a surface \cite{Vaittinen2014}. Therefore, before the calibration measurement, the adsorption was first saturated to a high level by flowing the sample gas for more than ten hours on a concentration level slightly higher than the first measurement level. The HF concentration was then dropped stepwise and adsorption was allowed to balance before taking a CEPAS signal reading. 


The stability and sensitivity of the PA system was investigated using Allan deviation. The CEPAS signal was measured with a high optical power of 950~mW at HF concentration of 92~ppt$_\textrm{v}$. The low stable HF concentration was generated by flushing the PA cell with dry ambient air for more than a week to reduce the residual desorption to a near constant level. The result is presented in Fig.~\ref{fig:AllanPlot}. The insete shows the data used for the Allan deviation plot except for the 1 and 2~s data points, calculated from the non-averaged data. The Allan deviation is used to extrapolate a noise equivalent concentration of the system from the SNR (SNR = 1) \cite{werle1993limits}. A noise equivalent concentration of 5.2~ppt$_\textrm{v}$ was reached in 1~s (one CEPAS reading), 2.5~ppt$_\textrm{v}$ was reached in one averaged gas exchange cycle (15~s), and 0.65~ppt$_\textrm{v}$ in 32 minutes. A normalized noise equivalent absorption (NNEA) coefficient of $2.7\times 10^{-10}$~W\,cm$^{-1}$Hz$^{-1/2}$ was reached with a sampling time of 1~s with an optical power of 950~mW. The stability was limited by drifting of the light source and fluctuations in the H$_2$O concentration of the sample gas. The observed H$_2$O fluctuations, in the order of 50~ppm$_\textrm{v}$ and in a time scale of few minutes, resulted in direct response in HF concentration since adsorption of HF is strongly dependent on the H$_2$O concentration.


In conclusion, a highly sensitive instrument for detecting hydrogen fluoride at 2.476~\textmu m was demonstrated using cantilever-enhanced photo-acoustic detection. Sub-ppt sensitivity is made possible by the combination of a sensitive detector, a very stable high power OPO and a strong rotational-vibrational transition of HF. The achieved noise equivalent concentrations (SNR=1) of 5.2~ppt$_\textrm{v}$ in 1~s and 0.65~ppt$_\textrm{v}$ in 32 min are, to our knowledge, both the lowest that have been reported for laser-based HF detection and for photo-acoustic spectroscopy, as well as one of the few laser-based gas analyzers capable of sub-ppt level trace gas detection \cite{Galli2011,Galli2016}. The resulting NNEA coefficient (1$\sigma$) of $2.7\times 10^{-10}$~W\,cm$^{-1}$Hz$^{-1/2}$ is also comparable to the best-reported NNEAs for CEPAS \cite{Peltola2015,Spagnolo2012}. The high sensitivity of the instrument is not limited to laboratory use, but the availability of commercial moderate power DFB lasers at the operation wavelength \citep{Craig2015} also allow the development of field-deployable sub-ppb level HF sensors. 

\textbf{Funding.} CHEMS doctoral program of the University of Helsinki; the Finnish Cultural Foundation; the Academy of Finland (grant numbers \#257479 and \#294752); the Finnish Funding Agency of Technology and Innovation (TEKES, grant number \#498/31/2015).


\bibliography{ArticleBibliography}

\begin{thebibliography}{10}
\newcommand{\enquote}[1]{``#1''}

\bibitem{NIOSH2007}
{National Institute for Occupational Safety}, \emph{NIOSH pocket guide to
  chemical hazards} (DIANE Publishing, 2007).

\bibitem{Duchatelet2010}
P.~Duchatelet, P.~Demoulin, F.~Hase, R.~Ruhnke, W.~Feng, M.~Chipperfield,
  P.~Bernath, C.~Boone, K.~Walker, and E.~Mahieu, Journal of Geophysical
  Research: Atmospheres \textbf{115} (2010).

\bibitem{Walna2013}
B.~Walna, I.~Kurzyca, E.~Bednorz, and L.~Kolendowicz, Environmental Monitoring
  and Assessment \textbf{185}, 5497 (2013).

\bibitem{Lim2009}
S.~H. Lim, L.~Feng, J.~W. Kemling, C.~J. Musto, and K.~S. Suslick, Nature
  chemistry \textbf{1}, 562 (2009).

\bibitem{Mertens2004}
J.~Mertens, E.~Finot, M.-H. Nadal, V.~Eyraud, O.~Heintz, and E.~Bourillot,
  Sensors and Actuators B: Chemical \textbf{99}, 58 (2004).

\bibitem{Dressler2008}
M.~Wolff and H.~Harde, Infrared Physics \& Technology \textbf{41}, 283 (2000).

\bibitem{Frish2005}
M.~Frish, R.~Wainner, B.~Green, M.~Laderer, and M.~Allen, \enquote{Standoff gas
  leak detectors based on tunable diode laser absorption spectroscopy,} in
  \enquote{Optics East 2005,}  (International Society for Optics and Photonics,
  2005), pp. 60100D--60100D.

\bibitem{Craig2015}
I.~M. Craig, B.~D. Cannon, M.~S. Taubman, B.~E. Bernacki, R.~D. Stahl, J.~T.
  Schiffern, T.~L. Myers, and M.~C. Phillips, Applied Physics B \textbf{120},
  505 (2015).

\bibitem{Rothman2013}
L.~S. Rothman, I.~E. Gordon, Y.~Babikov, A.~Barbe, D.~Chris~Benner, P.~F.
  Bernath, M.~Birk, L.~Bizzocchi, V.~Boudon, L.~R. Brown, A.~Campargue,
  K.~Chance, E.~A. Cohen, L.~H. Coudert, V.~M. Devi, B.~J. Drouin, A.~Fayt,
  J.~M. Flaud, R.~R. Gamache, J.~J. Harrison, J.~M. Hartmann, C.~Hill, J.~T.
  Hodges, D.~Jacquemart, A.~Jolly, J.~Lamouroux, R.~J. Le~Roy, G.~Li, D.~A.
  Long, O.~M. Lyulin, C.~J. Mackie, S.~T. Massie, S.~Mikhailenko, H.~S.~P.
  M\"{u}ller, O.~V. Naumenko, A.~V. Nikitin, J.~Orphal, V.~Perevalov,
  A.~Perrin, E.~R. Polovtseva, C.~Richard, M.~A.~H. Smith, E.~Starikova,
  K.~Sung, S.~Tashkun, J.~Tennyson, G.~C. Toon, V.~G. Tyuterev, and G.~Wagner,
  Journal of Quantitative Spectroscopy and Radiative Transfer \textbf{130}, 4
  (2013).

\bibitem{Tittel2003}
F.~K. Tittel, D.~Richter, and A.~Fried, \emph{Mid-infrared laser applications
  in spectroscopy} (Springer, 2003), pp. 458--529.

\bibitem{Vainio2016}
M.~Vainio and L.~Halonen, Phys. Chem. Chem. Phys. \textbf{18}, 4266 (20160).

\bibitem{Peltola2015}
J.~Peltola, T.~Hieta, and M.~Vainio, Opt Lett \textbf{40}, 2933 (2015).

\bibitem{Koskinen2006}
V.~Koskinen, J.~Fonsen, J.~Kauppinen, and I.~Kauppinen, Vibrational
  Spectroscopy \textbf{42}, 239 (2006).

\bibitem{Koskinen2007}
V.~Koskinen, J.~Fonsen, K.~Roth, and J.~Kauppinen, Applied Physics B: Lasers \&
  Optics \textbf{86}, 451 (2007).

\bibitem{Galli2011}
I.~Galli, S.~Bartalini, S.~Borri, P.~Cancio, D.~Mazzotti, P.~De~Natale, and
  G.~Giusfredi, Phys Rev Lett \textbf{107}, 270802 (2011).

\bibitem{Galli2016}
I.~Galli, S.~Bartalini, R.~Ballerini, M.~Barucci, P.~Cancio, M.~De~Pas,
  G.~Giusfredi, D.~Mazzotti, N.~Akikusa, and P.~De~Natale, Optica \textbf{3},
  385 (2016).

\bibitem{Vainio2008a}
M.~Vainio, J.~Peltola, S.~Persijn, F.~J. Harren, and L.~Halonen, Optics express
  \textbf{16}, 11141 (2008).

\bibitem{Vainio2008b}
M.~Vainio, J.~Peltola, S.~Persijn, F.~J.~M. Harren, and L.~Halonen, Applied
  Physics B \textbf{94}, 411 (2008).

\bibitem{Peltola2013}
J.~Peltola, M.~Vainio, T.~Hieta, J.~Uotila, S.~Sinisalo, M.~Mets\"{a}l\"{a},
  M.~Siltanen, and L.~Halonen, Opt. Express \textbf{21}, 10240 (2013).

\bibitem{Phillips2010}
C.~R. Phillips and M.~M. Fejer, J. Opt. Soc. Am. B \textbf{27}, 2687 (2010).

\bibitem{Westberg2012}
J.~Westberg, J.~Wang, and O.~Axner, Journal of Quantitative Spectroscopy and
  Radiative Transfer \textbf{113}, 2049 (2012).

\bibitem{DuBurck2004}
F.~Du~Burck and O.~Lopez, Measurement Science and Technology \textbf{15}, 1327
  (2004).

\bibitem{Sun1992}
H.~Sun and E.~Whittaker, Applied Optics \textbf{31}, 4998 (1992).

\bibitem{Vaittinen2014}
O.~Vaittinen, M.~Metsälä, S.~Persijn, M.~Vainio, and L.~Halonen, Applied
  Physics B \textbf{115}, 185 (2014).

\bibitem{werle1993limits}
P.~Werle, R.~M{\"u}cke, and F.~Slemr, Applied Physics B: Lasers and Optics
  \textbf{57}, 131 (1993).

\bibitem{Spagnolo2012}
V.~Spagnolo, P.~Patimisco, S.~Borri, G.~Scamarcio, B.~E. Bernacki, and
  J.~Kriesel, Opt. Lett. \textbf{37}, 4461 (2012).

\end{thebibliography}

 


\end{document}